\def\thisversion{20 August 2019}
\documentclass[aps,prd,twoside,twocolumn,nofootinbib]{revtex4-1}

\usepackage{amsmath,amssymb}
\usepackage{graphicx}
\usepackage{newtxtext,newtxmath}

\newcommand*{\qqtext}[1]{\qquad\text{#1}\qquad}

\allowdisplaybreaks

\begin{document}

\title{Model-independent form-factor constraints for electromagnetic spin-1 currents}%

\author{Helmut~Haberzettl}
 \email{helmut.haberzettl@gwu.edu}
 \affiliation{Institute for Nuclear Studies and Department of Physics, The George Washington University, Washington, DC 20052}

\date{\thisversion}

\begin{abstract}
Using local gauge invariance in the form of the Ward-Takahashi identity and
the fact that properly constructed current operators must be free of
kinematic singularities, it is shown that the magnetic moment $\mu$ and the
quadrupole moment $Q$ of an elementary spin-1 particle with mass $m$ and
charge $e$ are related by $2 m\mu + m^2 Q = e$, thus constraining the
normalizations of the Sachs form factors. This relation holds true as a
matter of course at the tree level in the standard model, but we prove it
remains true in general for dressed spin-1 states derived from elementary
fields. General expressions for spin-1 propagators and currents with
arbitrary hadronic dressing are given showing the result to be independent of
any dressing effect or model approach.\\[2ex]
DOI: 10.1103/PhysRevD.100.036008
\end{abstract}

\maketitle


\section{Introduction}

The electromagnetic structure of a massive spin-1 particle has been discussed
for some time (see
Refs.~\cite{LY1962,Lee1965,Aronson1969,KimTsai1973,Hagiwara1987,BH92,CBCT} and
references therein). The early work of Lee and Yang~\cite{LY1962} shows that at
the tree level, the particle's magnetic moment $\mu$ and the quadrupole moment
$Q$ are given by ($\hbar=c=1$) $\mu =e(1+\kappa)/2m$ and $Q=-e\kappa/m^2$ in
terms of one common constant $\kappa$. Although usually not written in this
manner, this correlation may also be expressed independent of $\kappa$ as
\begin{equation}
  2 m\mu + m^2 Q = e~,
  \label{eq:muQcorrelation}
\end{equation}
where $m$ is the mass and $e$ the charge. This relation is also true for the
canonical moments of the $W^\pm$ gauge boson in electroweak gauge theory at the
tree level where $\mu=e/m$ and $Q=-e/m^2$~\cite{KimTsai1973}, which corresponds
to putting $\kappa=1$ in the Lee-Yang result. The same expressions have also
been obtained by Brodsky and Hiller~\cite{BH92} in the strong binding limit
based on a generalization of the Gerasimov-Drell-Hearn sum
rule~\cite{Ger1965,DH1966}. The experimental value of $\mu$ for $W^\pm$, in
particular, of the DELPHI Collaboration~\cite{DELPHI}, quoted as the most
recent one by PDG~\cite{PDG2018}, is also compatible with this standard-model
result.

A more general electromagnetic structure allowing for the quadrupole moment to
be independent of charge and magnetic moment was considered in
Refs.~\cite{Aronson1969,KimTsai1973,Hagiwara1987,BH92,CBCT,LD1961,Jones1962,ACG1980}
(see also references therein), thus exploiting the full multipole degrees of
freedom of a spin-1 object. With the usual parametrization $\mu =
e(1+\kappa+\lambda)/2m$ and $Q= -e(\kappa-\lambda)/m^2$~\cite{KimTsai1973},
resulting in
\begin{equation}
  2 m\mu + m^2 Q = e\left(1+2\lambda\right)~,
  \label{eq:muQcorrelationmod}
\end{equation}
the value of $\lambda$ indicates the degree of independence of $\mu$ and $Q$.
The results tabulated in Ref.~\cite{CBCT} obtained for the $\rho$ meson by
various authors providing independent model determinations of $\mu$ and $Q$
correspond to values of $\lambda$ ranging from $0.1$ to about $0.5$, at
variance with the simple correlation (\ref{eq:muQcorrelation}).

In Sec.~\ref{sec:spin1}, we consider here the ramifications of imposing local
gauge invariance on the structure of the electromagnetic current operator of a
spin-1 particle based on an elementary field, and we will show in a
model-independent manner that Eq.~(\ref{eq:muQcorrelation}) is strictly valid
(i.e., $\lambda=0$) simply based on demanding a nonsingular current operator
that must satisfy the Ward-Takahashi identity~\cite{Ward50,Takahashi57,PSbook}
as a necessary and sufficient condition for \textit{local} gauge invariance.
Concluding remarks are provided in Sec.~\ref{sec:conclusions}.

\section{Spin-1 Current}\label{sec:spin1}

\begin{figure}[b!]
\centering
\includegraphics[width=1.0in,clip=]{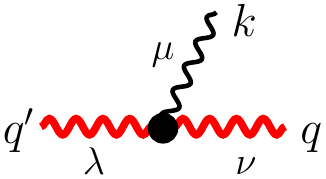}%
\caption{\label{fig:grho} Depiction of electromagnetic current vertex for the
$\rho$ meson, $\gamma(k) + \rho(q) \to \rho(q')$, with associated four-momenta
and Lorentz indices. (Time runs from right to left.)}
\end{figure}

To be locally gauge invariant, the spin-1 current $J^{\lambda\mu\nu}$ must
reproduce the Ward-Takahashi identity (WTI) of the form
\begin{equation}
k_\mu J^{\lambda\mu\nu}(q',q) \stackrel{!}{=} e \left[P^{-1}(q')-P^{-1}(q)\right]^{\lambda\nu}~,
\label{eq:WTI}
\end{equation}
where $P^{\lambda\nu}(q)$ is the propagator of the elementary spin-1 particle
with four-momentum $q$ and $k=q'-q$ is the (incoming) photon four-momentum (see
Fig.~\ref{fig:grho}). We emphasize here that except for the charge parameter
$e$, the right-hand side of the WTI contains no additional information about
the particle's electromagnetic structure. Moreover, the WTI is an
\textit{off-shell} relation at the operator level that necessarily requires a
commensurate off-shell structure for the associated current. The WTI must be
true irrespective of whether the spin-1 particle is a stable particle or a
resonance with nonzero width. It also must be true independent of the hadronic
gauge one chooses for, in general, the spin-1 propagator will be gauge
dependent~\cite{PSbook}. This gauge dependence will drop out when considering
physical matrix elements, however, to be consistent, it must be carried through
at all intermediate steps.

%
%

As usual, we  assume here the spin-1 particle to be stable, described by a
propagator $P^{\lambda\nu}(q)$ that has a physical pole with unit residue at a
real squared four-momentum $q^2=m^2$. [More general expressions will be
discussed at the end of this note, in Eqs.~(\ref{eq:Pdressed}) and
(\ref{eq:Jdressed}).] For a stable particle, the on-shell matrix element of the
inverse propagator vanishes, which will make the right-hand side of the WTI
(\ref{eq:WTI}) vanish for $q'^2=q^2=m^2$, thus indicating a gauge-invariant
conserved current.

The electromagnetic spin-1 current operator with form factors is usually
written as (see, e.g., Refs.~\cite{BH92,ACG1980})
\begin{align}
  J_0^{\lambda\mu\nu}(q',q) &= - e G_1(k^2) (q'+q)^\mu g^{\lambda\nu}
  \nonumber\\
  &\quad\mbox{}
  -eG_2(k^2) \left(k^\lambda g^{\mu\nu} - g^{\lambda\mu}k^\nu\right)
  \nonumber\\
  &\quad\mbox{}
  + eG_3(k^2)  (q'+q)^\mu \frac{k^\lambda k^\nu}{2m^2} ~,
  \label{eq:J0}
\end{align}
where the form factors $G_1$, $G_2$, and $G_3$ are related to the charge,
magnetic, and quadrupole form factors. This current ansatz comprises the most
general Lorentz structure available for on-shell matrix elements imposing
time-reversal invariance and current conservation. The four-momenta and Lorentz
indices appearing here are defined in Fig.~\ref{fig:grho} where the (charged)
$\rho$ meson is used as a generic template for an elementary spin-1 particle.

The Sachs form factors $G_C(k^2)$, $G_M(k^2)$, and $G_Q(k^2)$ describing
charge, magnetic moment, and quadrupole moment, respectively,
by~\cite{BH92,ACG1980}
\begin{equation}
\begin{pmatrix}
    G_1 \\ G_2 \\ G_3
  \end{pmatrix}
  =
  \begin{pmatrix}
    1  &  0& -\frac{2}{3}\eta
    \\
  0 &1 & 0
  \\
  -\frac{1}{1+\eta} & \frac{1}{1+\eta} &\frac{3+2\eta}{3+3\eta}
  \end{pmatrix}
  \begin{pmatrix}
    G_C \\ G_M \\ G_Q
  \end{pmatrix}~,
\label{eq:GGrelations}
\end{equation}
where $\eta=-k^2/4m^2$. Their normalizations are given by
\begin{subequations}\label{eq:GCG1etc}
\begin{align}
  eG_C(0) &= e  && \text{(charge ~ $e$)}~,
  \\
  eG_M(0) &= 2m \mu  &&\text{(magnetic moment ~ $\mu$)}~,
  \label{eq:G_M}
  \\
  eG_Q (0) & = m^2 Q   &&\text{(quadrupole moment ~ $Q$)}~,
\end{align}
\end{subequations}
which introduce the three electromagnetic multipole moments of the spin-1
particle. The corresponding normalizations of the form factors $G_i$
($i=1,2,3$) then are found as
\begin{subequations}\label{eq:G1G2G3}
\begin{align}
  G_1(0) &= G_C(0)=1~,
  \label{eq:G_1}
  \\
  G_2(0) &= G_M(0)= \frac{2m}{e} \mu~,
  \label{eq:G_2}
  \\
  G_3(0) &=-G_C(0) + G_M(0) +G_Q(0)
  \nonumber\\
  &=  -1+\frac{2m}{e}\mu+\frac{m^2}{e}Q~.
  \label{eq:G_3}
\end{align}
\end{subequations}
It is evident here in the last equation that $G_3(0)=0$ is equivalent to the
validity of Eq.~(\ref{eq:muQcorrelation}) and, indeed, we will show here that
the vanishing of $G_3(0)$ is a necessary condition for a well-defined current
that satisfies the WTI (\ref{eq:WTI}).

The four-divergence of the current (\ref{eq:J0}),
\begin{equation}
  k_\mu J_0^{\lambda\mu\nu} = e(q'^2-q^2)\left[-G_1(k^2) g^{\lambda\nu} +G_3(k^2) \frac{k^\lambda k^\nu}{2m^2} \right]~,
  \label{eq:kmuJ0}
\end{equation}
vanishes for $q'^2=q^2=m^2$ and thus indeed provides a conserved current.
However, this is not the correct form of the WTI for an elementary particle.
Clearly, to reproduce the WTI of the generic form (\ref{eq:WTI}), one must be
able to separate the four-divergence expression into a difference of two terms,
individually depending on $q'$ and $q$, respectively, without any $k^2$
dependence. This is simply not possible with form factors depending on $k^2$.

To resolve the discrepancy, one must move the electromagnetic form factors to
manifestly transverse terms, without changing the on-shell limit, similar to
the treatment of currents for spin-0 and spin-1/2 in Ref.~\cite{HH19}. To this
end, we may add an \textit{off-shell} term to the current (\ref{eq:J0})
according to
\begin{equation}
J_1^{\lambda\mu\nu}=J_0^{\lambda\mu\nu}
+ e k^\mu (q'^2-q^2)
\left(\frac{G_1-1}{k^2}  g^{\lambda\nu}
-\frac{G_3}{k^2}\, \frac{k^\lambda k^\nu}{2m^2}\right)
\label{eq:J01}
\end{equation}
that clearly is irrelevant for any physical matrix element and thus will not
change the electromagnetic form-factor content of the current as defined by
Eq.~(\ref{eq:J0}). However, this modification  is absolutely essential for
considerations of local gauge invariance in view of the fact that the
Ward-Takahashi identity itself is an off-shell relation. For the modified
current,
\begin{align}
  J_1^{\lambda\mu\nu}(q',q)
  &=
  -e (q'+q)^\mu g^{\lambda\nu}
  -e G_2 (k^\lambda g^{\nu\mu} - k^\nu g^{\mu\lambda})
  \nonumber\\[1ex]
  &\quad\mbox{}
  -  e\left(\frac{G_1-1}{k^2}  g^{\lambda\nu}-\frac{G_3}{k^2} \frac{k^\lambda k^\nu}{2m^2}\right)
  \nonumber\\
  &\qquad \mbox{}
  \times
    \left[ (q'+q)^\mu k^2 -k^\mu (q'^2-q^2) \right]~,
   \label{eq:J02}
\end{align}
the form-factor dependence does not appear in the four-divergence,
\begin{equation}
k_\mu J_1^{\lambda\mu\nu}(q',q) = -g^{\lambda\nu} e\left[(q'^2-m^2)-(q^2-m^2)\right]~,
\label{eq:WTI0}
\end{equation}
which has the correct structure of the WTI (\ref{eq:WTI}) and vanishes for
on-shell hadrons.

It should be emphasized here that the additional off-shell term in
Eq.~(\ref{eq:J01}) is unique because the resulting current expression
(\ref{eq:J02}) is comprised of the only three linearly-independent,
time-reversal-invariant \textit{transverse} operators available for spin 1 that
survive on shell.\footnote{There exists a fourth independent transverse
     operator, but since its on-shell matrix element vanishes, its coefficient
     function cannot be associated with a physical form factor and it can be
     put to zero without lack of generality. }
In other words, given $J^{\lambda\mu\nu}_0$ of Eq.~(\ref{eq:J0}), one cannot
construct an alternative subtraction curent that reproduces the WTI.

While the form (\ref{eq:WTI0}) of the WTI is only true for stable particles,
without any explicit hadronic dressing effects, it is sufficient for the
present purpose for it illustrates the basic mechanism how the dependence on
electromagnetic form factors is eliminated from the WTI.

The assertion that Eq.~(\ref{eq:muQcorrelation}) is true in general now simply
follows from noting that the operator structure of an electromagnetic current
must be free of kinematic singularities. We may demand, therefore, that the
additional current in Eq.~(\ref{eq:J01}) and thus the transverse term in the
modified current (\ref{eq:J02}) be well defined for all values of $q'$ and $q$.
In particular, it may not have singularities at the photon point, $k^2=0$,
which immediately provides the necessary conditions
\begin{equation}
  G_1(0) =1
  \qqtext{and}
  G_3(0) = 0
  \label{eq:G1G3}
\end{equation}
to make $(G_1-1)/k^2$ and $G_3/k^2$ well behaved. The first condition is
trivially true because of the normalization (\ref{eq:G_1}). The second
condition then makes the right-hand side of Eq.~(\ref{eq:G_3}) vanish, which is
equivalent to (\ref{eq:muQcorrelation}), and thus proves the point that the
validity of Eq.~(\ref{eq:muQcorrelation}) is not limited to the assumptions of
the original Lee-Yang approach~\cite{LY1962}, but remains true in general.

\subsection{Fully dressed spin-1 current}

We complete the presentation here by showing that even allowing for arbitrary
dressing effects will not alter this conclusion.

Without going into details, the most general fully dressed spin-1 propagator
may be written in terms of two (in general, complex) scalar dressing functions
as
\begin{equation}
  P^{\lambda\nu}(q) = \frac{-g^{\lambda\nu}
  + \frac{q^\lambda q^\nu}{m^2}N(q^2)}{q^2-m^2 - \Sigma(q^2)}
   ~.
   \label{eq:Pdressed}
\end{equation}
Here, $N(q^2)$ is a gauge-dependent function that is irrelevant for physical
matrix elements. The selfenergy function $\Sigma(q^2)$, on the other hand,
determines all physically relevant dressing effects. To make $m$ the physical
mass, it is assumed here that the selfenergy vanishes at $q^2=m^2$ , but this
can be arranged easily. The inverse of the propagator, as it appears in the
generic WTI (\ref{eq:WTI}), reads
\begin{equation}
\left(P^{-1}(q)\right)^{\lambda\nu} = -g^{\lambda\nu} D(q^2) + q^\lambda q^\nu C(q^2)
\label{eq:Pinvdressed}
\end{equation}
where
\begin{equation}
  D(q^2) =q^2-m^2 - \Sigma(q^2)
\end{equation}
is a short-hand notation for the denominator of the propagator
(\ref{eq:Pdressed}). The function $C(q^2)$ contains $N(q^2)$ and thus is gauge
dependent; its details can easily be worked out by explicitly constructing the
inverse (\ref{eq:Pinvdressed}), but since they are not relevant, they will be
omitted here.

The fully dressed current compatible with the propagator (\ref{eq:Pdressed})
then is obtained by applying the gauge derivative \cite{HH19,HH97} to the
inverse propagator (\ref{eq:Pinvdressed}) resulting in
\begin{align}
J^{\lambda\mu\nu}(q',q) &=J_1^{\lambda\mu\nu}(q',q) \frac{D(q'^2)-D(q^2)}{q'^2-q^2}
+ J_{\text{gauge}}^{\lambda\mu\nu}(q',q)~,
\label{eq:Jdressed}
\end{align}
with a gauge-dependent current piece that reads
\begin{align}
J_{\text{gauge}}^{\lambda\mu\nu}(q',q) &=eq'^\lambda  g^{\mu\nu} C(q'^2) + e g^{\lambda\mu} q^\nu C(q^2)
\nonumber\\[1ex]
&\quad\mbox{}
+e q'^\lambda (q'+q)^\mu  q^\nu \frac{C(q'^2)-C(q^2)}{q'^2-q^2}~,
\label{eq:Jgauge}
\end{align}
whose on-shell matrix elements vanish. The $0/0$ situations arising here at
$q'^2=q^2$ from the finite-difference derivatives of the denominator function
$D$ in (\ref{eq:Jdressed}) and of the function $C$ in (\ref{eq:Jgauge}) are
well behaved and nonsingular. For a stable particle, in particular, the
on-shell value of the finite-difference derivative of $D$ is directly related
to the unit residue of the propagator and thus unity as well. Hence, the
on-shell matrix elements of the current $J^{\lambda\mu\nu}$ with full hadronic
dressing, of the modified undressed current $J_1^{\lambda\mu\nu}$, and of the
usual current expression $J_0^{\lambda\mu\nu}$ of Eq.~(\ref{eq:J0}) are
identical. The normalizations in Eqs.~(\ref{eq:GCG1etc}) and (\ref{eq:G1G2G3}),
therefore, are not affected by hadronic dressing.

Evaluating now the four-divergences of the gauge-dependent current
contribution,
\begin{equation}
k_\mu J_{\text{gauge}}^{\lambda\mu\nu}(q',q)=e\left[ q'^\lambda q'^\nu C(q'^2)-q^\lambda q^\nu C(q^2)\right]~,
\end{equation}
and of the entire dressed current,
\begin{align}
k_\mu J^{\lambda\mu\nu}(q',q)
&= -g^{\lambda\nu} e\left[D(q'^2)
-D(q^2)\right]
+ k_\mu J_{\text{gauge}}^{\lambda\mu\nu}(q',q)~,
\label{eq:Jfull4div}
\end{align}
we indeed obtain the WTI (\ref{eq:WTI}) in terms of the fully dressed inverse
propagator (\ref{eq:Pinvdressed}). The dressed current (\ref{eq:Jdressed}),
therefore, is locally gauge invariant. Moreover, for a stable spin-1 particle,
the physical on-shell matrix element of the four-divergence
(\ref{eq:Jfull4div}) vanishes, thus providing a conserved current.

All electromagnetic form factors appear here only in $J_1^{\lambda\mu\nu}$ in
Eq.~(\ref{eq:Jdressed}) in manifestly transverse contribution, as detailed in
Eq.~(\ref{eq:J02}). Hence, the demand that these contributions should be well
behaved and free of kinematic singularities carries over directly to the
present case with full hadronic dressing. The conditions (\ref{eq:G1G3}),
therefore, are valid here as well, independent of the details of dressing
effects.

\section{Conclusion and Discussion}\label{sec:conclusions}

We may thus conclude that the relationship (\ref{eq:muQcorrelation}) linking
the three multipole moments of an elementary spin-1 particle holds true in
general and that it is model independent. While this correlation is trivially
satisfied by the canonical moment values (i.e., $\mu=e/m$, $Q=-e/m^2$)
discussed in the first paragraph of the Introduction, the relationship as such
does not make any demand on individual values other than that they must be
linked to satisfy (\ref{eq:muQcorrelation}). This correlation is important on
general theoretical grounds because it reduces the multipole degrees of
freedom. Moreover, since it imposes restrictions on approximations in model
treatments of the moments, it will also allow for more realistic assessments of
the reliability of various approaches. Other than possessing spin 1, the
derivation makes no special demands on the nature of the particle as long as
its description is based on an elementary field. It therefore applies to the
$W^\pm$ gauge bosons of electroweak theory as well as to strongly interacting
spin-1 particles like the $\rho$ meson, etc.

However, the present considerations do \textit{not} apply to hadronic spin-1
bound states like the deuteron because the requirement of the
WTI~(\ref{eq:WTI}) as a necessary and sufficient statement of local gauge
invariance only applies to elementary particles. For hadronic bound states, in
principle, their electromagnetic structures can be described microscopically in
terms of how their (observable) hadronic constituents couple to the
electromagnetic field. Comprehensive gauge-invariance considerations for
hadronic spin-1 bound states like the deuteron, therefore, would need to
consider also the possibility of asymptotically free constituent particles,
including their final-state interactions. A somewhat simplified (incomplete)
description along such lines can be found in Ref.~\cite{LD1961}. Hence,
utilizing the current (\ref{eq:J0}) for the
deuteron~\cite{LD1961,Jones1962,ACG1980} provides an \textit{effective}
description of its electromagnetic spin-1 properties, with a conserved current
because the four-divergence (\ref{eq:kmuJ0}) vanishes on shell, however, there
is no associated ``deuteron propagator'' to satisfy the WTI (\ref{eq:WTI})
independent of its effective electromagnetic properties.

Finally, we mention without further discussion that in the elementary-particle
case, the respective expressions for the dressed propagator,
Eq.~(\ref{eq:Pdressed}), and the dressed current, Eq.~(\ref{eq:Jdressed}),
remain valid even if the spin-1 particle is a resonance, with nonzero width
described by the imaginary part of the dressing function $\Sigma$. The mass $m$
and the moments $\mu$ and $Q$ then are parameters tied together by the
normalizations (\ref{eq:G1G2G3}), but they will not necessarily retain their
usual physical meanings if the width is too large.


\acknowledgments

The author acknowledges partial support by the U.S. Department of Energy,
Office of Science, Office of Nuclear Physics, under Award Number
\mbox{DE}-\mbox{SC0016582}.


\end{document}